\documentclass[runningheads]{llncs}
\usepackage{siunitx}
\usepackage{wrapfig}
\usepackage{graphicx}
\usepackage{subfigure}
\usepackage{dsfont}
\usepackage{caption}
\usepackage{color}
\usepackage{pgfplots}
\usepackage{float}
\usepackage{amsfonts}

\usepackage{bm}
\usepackage[misc]{ifsym}
\usepackage{amssymb}
\usepackage{bm}
\usepackage{booktabs} % For formal tables
\usepackage{tikz,xcolor,caption}
\usetikzlibrary{positioning}
\usetikzlibrary{automata}
\usetikzlibrary{positioning,automata,shadows}
\usepackage{tikz}
\usetikzlibrary{arrows}
\usetikzlibrary{automata,positioning}
\usetikzlibrary{decorations.markings,calc}
\usetikzlibrary{shadows}
\tikzset{
    overlaid/.style={double copy shadow={shadow xshift=-1ex,shadow yshift=1.5ex},fill=white,draw=black,thick,minimum height = 1.5cm,minimum width=1cm,text width = 1.5cm, align=center},
}
\tikzset{
    process/.style={
        inner sep=0,
        text width=1.5cm, draw,
        minimum height=1.2cm,
        text centered,
        },
    process1/.style={
        text width=1.5cm, draw,
        minimum height=0.8cm,
        text centered,
        },
    description/.style={
        text centered,
        text width=10cm,
    },
    myarrow/.style={
        postaction={
            decorate, decoration={
                markings,mark=at position #1 with {\arrow{Stealth};
                }
            }
        }
    },
}
 \tikzstyle{accepting}=[path picture={%
  \draw let
    \p1 = (path picture bounding box.east),
    \p2 = (path picture bounding box.center)
    in
      (\p2) circle (\x1 - \x2 - 2pt);
  }]
   \tikzset{test/.style={
    postaction={
        decorate,
        decoration={
            markings,
            mark=at position \pgfdecoratedpathlength-0.5pt with {\arrow[blue,line width=#1] {>}; },
            mark=between positions 0 and \pgfdecoratedpathlength-8pt step 0.5pt with {
                \pgfmathsetmacro\myval{multiply(divide(
                    \pgfkeysvalueof{/pgf/decoration/mark info/distance from start}, \pgfdecoratedpathlength),100)};
                \pgfsetfillcolor{blue!\myval!red};
                \pgfpathcircle{\pgfpointorigin}{#1};
                \pgfusepath{fill};}
}}}}

\usepackage{multirow}
\usepackage{xstring,calc}
\makeatletter
    \newif\if@restonecol
    \makeatother

    \usepackage[linesnumbered,ruled,vlined]{algorithm2e}%[ruled,vlined]{
    \usepackage{algpseudocode}
    \usepackage{amsmath}
\usetikzlibrary{calc}

\usepackage{array}
\newcolumntype{I}{!{\vrule width 0.7pt}}
\newlength\savedwidth

\newlength\savewidth

\usepackage{bbding}

\definecolor{mypink1}{rgb}{0, 0, 0}
\definecolor{liyt_red}{RGB}{254,117,109}
\definecolor{liyt_green}{RGB}{108,206,157}
\definecolor{liyt_blue}{RGB}{102,188,243}
\definecolor{liyt_gray}{RGB}{166,170,179}
\definecolor{bella}{RGB}{0, 0, 0}
\definecolor{Maroon}{rgb}{0,0,0}
\begin{document}
\title{An Effective Algorithm for Learning Single Occurrence Regular Expressions with Interleaving}
\titlerunning{An Effective Algorithm for Learning SOIREs with Interleaving}
\author{Yeting Li\inst{1,2}  \and Haiming Chen\inst{1} \and Xiaolan Zhang\inst{1,2} \and Lingqi Zhang\inst{3}}
\authorrunning{Y. Li et al.}

\institute{State Key Laboratory of Computer Science, Institute of Software, Chinese Academy of Sciences, Beijing 100190, China\\\email{\{liyt,chm,zhangxl;\}@ios.ac.cn}
\and University of Chinese Academy of Sciences, Beijing, China
\and Beijing University of Technology Beijing, China\\\email{zhanglingqisteve@gmail.com}
}
\maketitle

\begin{abstract}
The advantages offered by the presence of a schema are numerous.
However, many XML documents in practice are not accompanied by a (valid) schema, making schema inference an attractive research problem.
The fundamental task in XML schema learning is inferring restricted subclasses of regular expressions.
Most previous work either lacks support for interleaving or only has limited support for interleaving.
%lacks support of interleaving. Hence, the support of interleaving in existing works is very limited.
%Most of the previous work in this area either lacked support for interleaving or limited support for interlacing.
%based on the analysis of large-scale real data,
In this paper, we first propose a new subclass  \textit{Single Occurrence Regular Expressions with Interleaving} (SOIRE), which has unrestricted support for interleaving. Then, based on \textit{single occurrence automaton} and \textit{maximum independent set}, we propose an algorithm $i$SOIRE to infer SOIREs.
Finally, we further conduct a series of experiments on real datasets to evaluate the effectiveness of our work, comparing with both ongoing learning algorithms in academia and industrial tools in real-world.
The results reveal the practicability of SOIRE and the effectiveness of $i$SOIRE, showing the high preciseness and conciseness of our work.
\end{abstract}


\section{Introduction}

%xml schema 很重要，作用如下
XML schemas have always played a crucial role in XML management.
The presence of a schema for XML documents has many advantages, such as for query processing and optimization, development of database applications, data integration and exchange~\cite{DBLP:conf/icde/ColazzoGS11,DBLP:journals/pvldb/WangHZS15,DBLP:journals/tods/MartensNNS17,DBLP:journals/is/GallinucciGR18}.
%但是现实中缺失了很多schema,所以有必要推到schema
However, many XML documents in practice are not accompanied by a (valid) schema~\cite{DBLP:journals/www/BarbosaMV05,DBLP:conf/www/MignetBV03,DBLP:journals/tods/MartensNSB06,DBLP:conf/webdb/BexNB04,DBLP:conf/webdb/Sahuguet00a,DBLP:journals/ws/GrijzenhoutM13}, making schema inference an attractive research problem~\cite{Baazizi2019,DBLP:journals/tweb/BexGNV10,DBLP:conf/vldb/BexNST06,DBLP:journals/mst/FreydenbergerK15,DBLP:journals/datamine/GarofalakisGRSS03,DBLP:conf/pakdd/ZhangLCDC18,DBLP:journals/corr/CiucanuS13,DBLP:conf/adma/LiMC18,DBLP:conf/er/LiZXMC18}.
%,LiZXMCER18
%研究schema推断还有其他应用
Studying schema inference also has several practical motivations.
Schema inference techniques may be extended to schema repairing techniques~\cite{DBLP:journals/ws/GrijzenhoutM13}.
Besides, schema inference is also useful in situations where a schema is already available, such as in schema cleaning and dealing with noise~\cite{DBLP:conf/vldb/BexNST06}.

 The content models of XML schemas are defined by regular expressions, and previous research has shown that the essential task in schema learning is inferring regular expressions from a set of given samples~\cite{DBLP:conf/vldb/BexNV07,Baazizi2019,DBLP:journals/tweb/BexGNV10,DBLP:conf/vldb/BexNST06,DBLP:journals/mst/FreydenbergerK15,DBLP:journals/datamine/GarofalakisGRSS03,DBLP:conf/pakdd/ZhangLCDC18,DBLP:journals/corr/CiucanuS13,DBLP:conf/adma/LiMC18,DBLP:conf/er/LiZXMC18}. In fact, in some cases these learned regular expressions can directly be used as parts of the schema, and in other cases the inference of regular expressions is the most important component of the schema inference. Therefore, research on schema learning has focused on inferring regular expressions from a set of given samples.

%而我们关注的推断interleaving 或者Shuffled Languages,因为interleaving应用很广
We focus on learning regular expressions with \textit{interleaving} (\textit{shuffle}), denoted by RE(\&). Since RE(\&) are widely used in various areas of computer science~\cite{BerglundBB13},
including XML database systems~\cite{XSD,RELAXNG,DBLP:journals/tods/MartensNNS17}, complex event processing~\cite{DBLP:conf/icde/LiG15}, system verification~\cite{DBLP:conf/lics/BojanczykMSSD06,DBLP:journals/tcs/GargR92,DBLP:journals/cacm/Gischer81}, plan recognition~\cite{DBLP:conf/fusion/HogbergK10} and natural language processing~\cite{DBLP:conf/eacl/KuhlmannS09,DBLP:conf/acl/Nivre09}.

Inference of regular expressions from a set of given samples belongs to the problem of language learning. Gold proposed a classical language learning model (\textit{learning in the limit or explanatory learning}) and pointed out that the class of regular expressions could not be identifiable from positive samples only \cite{DBLP:journals/iandc/Gold67}. This means that no matter how many positive samples from the target language (i.e., the language to be learned) are provided, no algorithm can infer every target regular expression. Hence, researchers have turned to study  subclasses of regular expressions \cite{DBLP:journals/ipl/MinAC03,DBLP:conf/vldb/BexNV07,Baazizi2019,DBLP:journals/tweb/BexGNV10,DBLP:conf/vldb/BexNST06,DBLP:journals/mst/FreydenbergerK15,DBLP:journals/datamine/GarofalakisGRSS03,DBLP:conf/pakdd/ZhangLCDC18,DBLP:journals/corr/CiucanuS13,DBLP:conf/adma/LiMC18,DBLP:conf/er/LiZXMC18}.

Most existing subclasses of regular expressions for XML are defined on standard regular expressions, e.g.,~\cite{DBLP:journals/tweb/BexGNV10,DBLP:conf/vldb/BexNST06,DBLP:conf/webdb/BexNB04,Feng20,Martens2013Complexity} which were analyzed together in \cite{DBLP:conf/apweb/LiZPC16,DBLP:conf/ideas/LiCMDC18}. For single occurrence regular expressions (SOREs), in which each symbol occurs at most once and its subclass chain regular
expressions (CHAREs), Bex et al. proposed two inference algorithms \textit{RWR} and \textit{CRX}~\cite{DBLP:conf/vldb/BexNST06,Bex2010Inference}.
Freydenberger and K\"{o}tzing~\cite{DBLP:journals/mst/FreydenbergerK15} proposed more efficient algorithms \textit{Soa2Sore} and \textit{Soa2Chare} for the above mentioned
SOREs and CHAREs. Bex et al.~\cite{DBLP:journals/tweb/BexGNV10} also studied learning algorithms, based on the Hidden Markov Model, for the subclass of regular expressions ($k$-OREs) in which each alphabet symbol occurs at most $k$ times. Notice that none of the above subclasses support an important feature in XML, i.e., the interleaving.

There may be no order constraint among siblings in data-centric applications~\cite{DBLP:journals/mst/AbiteboulBV15}. In such cases the interleaving is necessary.
Here we list the more recent efforts on RE(\&) inference (see~\cite{DBLP:journals/corr/CiucanuS13,DBLP:conf/apweb/PengC15,DBLP:conf/pakdd/ZhangLCDC18,DBLP:conf/adma/LiMC18,DBLP:conf/er/LiZXMC18}). The aim of these approaches is to infer restricted subclasses of single occurrence RE(\&), in which each symbol occurs at most once, starting from a positive set of words.
Ciucanu and Staworko proposed two subclasses disjunctive multiplicity expression (DME) and disjunction-free multiplicity expression (ME)~\cite{DBLP:conf/webdb/BonevaCS13,DBLP:journals/corr/CiucanuS13} which support unordered concatenation, a weaker form of interleaving. The concatenation operator is disallowed in both formalisms and ME even uses no disjunction operator.
For example, $r_1=(a|b^+)\& c$ is a DME and $r_2=a\& b^* \& c^?$ is an ME. But $r_3=(a^+b^?)\& c^*$ and $r_4=a^*((b^*|c)\&d^*)$ do not satisfy both formalisms.
 The inference algorithm based on \textit{maximum clique} for DME was given in~\cite{DBLP:journals/corr/CiucanuS13}.
Li et al. provided an algorithm to learn DMEs from both positive and negative examples based on genetic algorithms and simplified candidate regions (SCRs)~\cite{DBLP:conf/dasfaa/LiDCC19}.
  When there is no order constraint among siblings, the relative orders within siblings are still important. Peng and Chen~\cite{DBLP:conf/apweb/PengC15} proposed a subclass SIRE
 using the grammar: $S::=T\&S|T$, $T::=\varepsilon |a|a^*|TT$. But it does not support the union operator.
 For example, $r_2$ and $r_3$ are SIREs but $r_1$ and $r_4$ are not. Besides, they presented an approximate algorithm to infer SIREs~\cite{DBLP:conf/apweb/PengC15}.
 Li et al.~\cite{DBLP:conf/adma/LiMC18} proposed a subclass ICRE
 using the grammar: \begin{align*}
& E:= F_1^{p_1}\cdot \ldots \cdot F_n^{p_n},&&(n\geq 0,p_i \in\{?,1\}),\\
& F_i:=D_1 \& \dots \& D_k, &&(i \in [1,n],  k\geq 1),\\
& D_j:= a_1^{mul_1}| \dots | a_m^{mul_m},&&(j\in [1,k], m\geq 1),~~~~~~~~
\end{align*} where $mul_o\in \{1,?,*,+\}$ and $a_o\in \Sigma$ for $o\in [1,m]$.
For example, $r_1$, $r_2$ and $r_4$ are ICREs but $r_3$ is not.
Besides, they presented an approximate algorithm to infer ICREs~\cite{DBLP:conf/adma/LiMC18}.
 Zhang et al.~\cite{DBLP:conf/pakdd/ZhangLCDC18} proposed a subclass called ICHARE considering interleaving. The inference algorithm is based on SOA and maximum independent set (MIS). However, components of interleaving are restricted to the \textit{extended strings} (ES) defined in \cite{DBLP:conf/pakdd/ZhangLCDC18}. For example, $r_2$ and $r_3$ are ICHAREs but $r_1$, $r_4$ and $r_5=a^?((b^+|c)d^*\&ef^?)$ are not.
Li et al.~\cite{DBLP:conf/er/LiZXMC18} proposed a practical subclass called ESIRE and designed an inference algorithm GenESIRE to infer ESIREs.
For example, $r_1$, $r_2$, $r_3$, $r_4$ and $r_5$ are ESIREs, but $r_6 = a^*b^?(fm^?\&c^?d|e(n|l)^?g\&h^?)(j^+|k)^?$ is not.
All of the above subclasses are restricted subclasses of single occurrence RE(\&).
As shown above, the support for interleaving in existing work is very limited.

\begin{figure}
\scriptsize
\begin{minipage}[b]{0.5\linewidth}
\subfigure[ME $\subset$ DME $\subset$ ICRE]{
\begin{tikzpicture}
\draw (-2.0,-1.7) rectangle (2.0,1.7) node[below left]{RE(\&)};
\draw (0,0) circle (0.5cm) node[text=black] {ME};
\draw (0,0) circle (1.0cm) node {};%{DME};
\draw (0,0) circle (1.5cm) node {};%[above] {ICRE};
\node  (dme) at (0,-0.7) {DME};
\node  (icre) at (0,-1.25) {ICRE};
\end{tikzpicture}
}
\end{minipage}\begin{minipage}[b]{0.5\linewidth}
\subfigure[ME $\subset$ SIRE $\subset$ ICHARE]{
\begin{tikzpicture}
\draw (-2.0,-1.7) rectangle (2.0,1.7) node[below left]{RE(\&)};
\draw (0,0) circle (0.5cm) node[text=black] {ME};
\draw (0,0) circle (1.0cm) node {};
\draw (0,0) circle (1.5cm) node {};
\node  (dme) at (0,-0.7) {SIRE};
\node  (icre) at (0,-1.25) {ICHARE};
\end{tikzpicture}
}
\end{minipage}
\begin{minipage}[b]{0.5\linewidth}
\subfigure[DME $\cap$ SIRE $=$ ME]{
\begin{tikzpicture}
\draw (-2.0,-1.7) rectangle (2.0,1.7) node[below left]{RE(\&)};

%\draw (-2,-1.5) rectangle (3.5,1.5) node[below left]{$U$};
\begin{scope} % start of clip scope
\clip (-0.75,0) circle (1cm);
\fill[gray] (0.75,0) circle (1cm);
\end{scope} % end of clip scope
\draw (-0.75,0) circle (1cm) node[left] {DME};
\draw (.75,0) circle (1cm) node[right] {SIRE};
\node  (me) at (0,-0.95) {ME};
\end{tikzpicture}
}
\end{minipage}\begin{minipage}[b]{0.5\linewidth}
\subfigure[ICRE $\subset$ ESIRE $\subset$ SOIRE $\subset$ RE(\&), ICHARE $\subset$ ESIRE $\subset$ SOIRE $\subset$ RE(\&)]{
\begin{tikzpicture}
\draw (-2.0,-1.7) rectangle (2.0,1.7) node[below left]{RE(\&)};
%\draw (-4.3,-2) rectangle (4.3,2) node[below left]{RE(\&)};
\draw (-0.48,0) circle (0.8cm) node[left] {\tiny{ICRE}};
\draw (0.47,0) circle (0.8cm) node[right] {};%{ICHARE};
\draw (0,0) circle (1.3cm) node[below] {};%{SOIRE};
\draw (0,0) circle (1.6cm) node[below] {};%{SOIRE};
\node  (icre) at (0,-1.4) {SOIRE};
\node  (icre) at (0,-1.0) {ESIRE};
\node  (ichare) at (0.8,0) {\tiny{ICHARE}};
\end{tikzpicture}
}
\end{minipage}
\caption{Relationships among ME, DME, SIRE, ICRE, ICHARE, ESIRE, SOIRE and RE(\&).}
\label{fig:rel}
\end{figure}
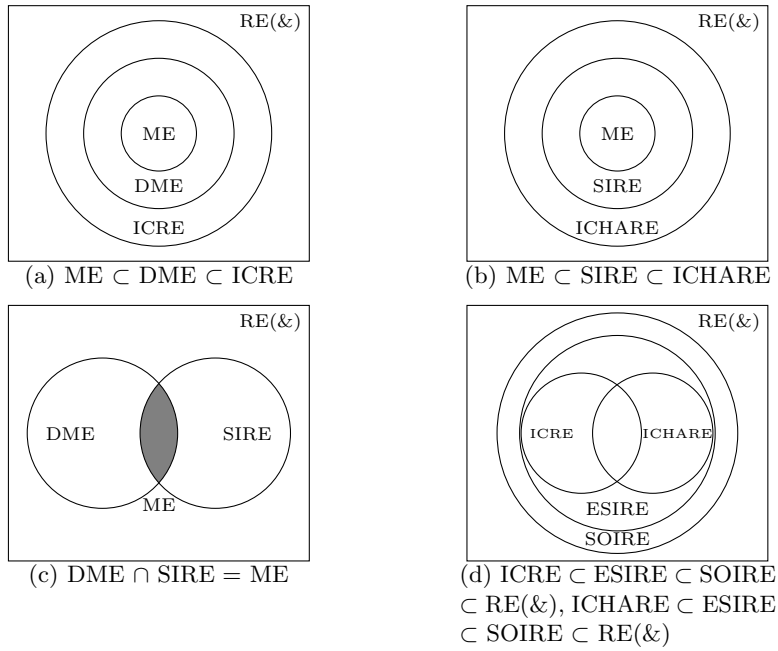

In this paper, based on the analysis of large-scale real data, we propose a new subclass of RE(\&), i.e., single occurrence RE(\&), called SOIRE.
The relationships among ME, DME, SIRE, ICRE, ICHARE, ESIRE, SOIRE and RE(\&) are shown in Figure~\ref{fig:rel}.
Among them, ME $\subset$ DME $\subset$ ICRE, ME $\subset$ SIRE $\subset$ ICHARE, DME $\cap$ SIRE $=$ ME, ICRE $\subset$ ESIRE $\subset$ SOIRE $\subset$ RE(\&) and ICHARE $\subset$ ESIRE $\subset$ SOIRE $\subset$ RE(\&).
For example, all of $r_1$, $r_2$, $r_3$, $r_4$, $r_5$ and $r_6$ are SOIREs.
It reveals that SOIRE is more powerful than the above subclasses since the latter are all subclasses of SOIRE, and especially SOIRE has unrestricted support for interleaving, which was never achieved by  existing work.
Then, we develop the corresponding learning algorithm, $i$SOIRE, to carry out SOIREs inference automatically.
The massive experimental results demonstrate the practicality of the proposed subclass as well as
the preciseness and conciseness of $i$SOIRE.

The main contributions of this paper are listed as follows.
\begin{itemize}
    \item We propose a new subclass SOIRE of RE(\&). SOIRE is more powerful than the existing subclasses and especially has unrestricted support for interleaving.
    \item Correspondingly, we design an inference algorithm $i$SOIRE which can learn SOIREs effectively based on single occurrence automaton (SOA) and maximum independent set (MIS).
    \item We conduct a series of experiments, comparing the performance of our algorithm with both ongoing learning algorithms in academia and industrial tools in real-world.
    The results reveal the practicability of SOIRE and the effectiveness of $i$SOIRE, showing the high preciseness and conciseness of our work.
\end{itemize}

\iffalse
This paper is organized as follows. Section~\ref{sec2} presents the basic definitions. Section $3$ gives the inference algorithm \textit{GenESIRE}. Section $4$ introduces the experiments. Conclusions are drawn in Section $5$.
\fi

The rest of this paper is organized as follows. Preliminaries are presented in Section~\ref{sec2}. Section~\ref{sec3} provides the learning algorithm. Then a series of experiments is presented in Section~\ref{sec4}. Finally we conclude this work in Section~\ref{sec5}.

\section{Preliminaries}\label{sec2}
\subsection{Definitions}
Let $\Sigma$ be a finite alphabet of symbols. The set of all words over $\Sigma$ is denoted by $\Sigma^*$. The empty word is denoted by $\varepsilon$. 
\begin{definition}
\textbf{Regular Expression with Interleaving.} A regular expression with interleaving over $\Sigma$ is defined inductively as follows: $\varepsilon$ or $a \in \Sigma$ is a regular expression, for regular expressions $r_1$ and $r_2$, the disjunction $r_1 | r_2$, the concatenation $r_1 \cdot r_2$, the interleaving $r_1\&r_2$, or the Kleene-Star $r_1^*$ is also a regular expression. $r^?$ and $r^+$ are abbreviations of $r|\varepsilon$ and $r \cdot r^*$, respectively. They are denoted as RE(\&).

\end{definition}
The size of a regular expression $r$, denoted by $|r|$, is the total number of symbols and operators occurred in $r$. The language $L(r)$ of a regular expression $r$ is defined as follows: $L(\varnothing)=\varnothing$; $L(\varepsilon)=\{\varepsilon\}$; $L(a)=\{a\}$; $L(r_1^*)=L(r_1)^*$; $L(r_1$$\cdot$$ r_2)=L(r_1)L(r_2)$; $L(r_1|r_2)=L(r_1)$$\cup$$L(r_2)$; $L(r_1$$\&$$r_2)=L(r_1)$$\&$$L(r_2)$.
%\end{definition}
Let $u=au'$ and $v=bv'$ where $a,b$$\in$$\Sigma$ and $u',v'$$ \in$$\Sigma^*$, then $u\&\varepsilon = \varepsilon\&u = \{u\}$ and $u\&v = a(u'\&v)$$\cup$ $b(u\&v')$. For example, $L(ab$$\&$$c)$ = $\{cab,acb,abc\}$.

\begin{definition}
\label{def2}
\textbf{Single Occurrence Regular Expressions with Interleaving (SOIRE).}
A regular expression with interleaving is SOIRE, in which each symbol occurs at most once.
\end{definition}

For instance, $r_1 =a^?(b^?c\&d^*(e|f)^?)$ is an SOIRE, but $r_2 = a^+b\&c^+b$ is not because $b$ appears twice.

\begin{definition}
\textbf{Single Occurrence Automaton (SOA)} \cite{DBLP:conf/vldb/BexNST06,DBLP:journals/mst/FreydenbergerK15} Let $\Sigma$ be a finite alphabet. $src$ and $snk$ are distinct symbols that do not occur in $\Sigma$. A single occurrence automaton (short: SOA) over $\Sigma$ is a finite directed graph $\mathcal{A} = (V,E)$ such that

\begin{enumerate}
  \item $src,snk \in V$, and $V\subseteq \Sigma \cup \{src,snk\}$;
  \item $src$ has only outgoing edges, $snk$ has only incoming edges and every node $v \in V$ lies on a path from $src$ to $snk$.
\end{enumerate}
\end{definition}

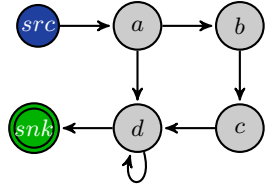
\begin{figure}[th]

\centering
\begin{tikzpicture}[->,>=stealth',shorten >=1pt,auto,node distance=0.7cm,
  thick,main node/.style={circle,fill=black!20,draw,minimum size=18pt}
]
\node[main node,inner sep=1pt,minimum size=0pt,fill={rgb:red,1;green,2;blue,5}] (src) {\textcolor{white}{\textbf{$src$}}}; %,initial
\node[main node] (a1) [right =of src] {$a$};
\node[main node] (a2) [right =of a1] {$b$};
\node[main node] (b1) [below =of a2] {$c$};
\node[main node] (b2) [left =of b1] {$d$};
\node[main node,accepting,inner sep=1pt,minimum size=0pt,fill=black!30!green](sink) [left =of b2] {\textcolor{white}{$snk$}};
\path[->,every node/.style={font=\sffamily\small}]
(src) edge   (a1)
(a1) edge   (a2)
(a1) edge   (b2)
(a2) edge   (b1)
(b1) edge (b2)
(b2) edge (sink)
edge [loop below]  ();
\end{tikzpicture}
\caption{Example SOA $\bm{\mathcal{A}}$ for $\bm{r = a(bc)^?d^+$}.}
\label{fig:fig2}
\end{figure}

For example, the SOA $\mathcal{A}$ for $r = a(bc)^?d^+$ is shown in Figure~\ref{fig:fig2}.
A \textbf{generalized single occurrence automaton (generalized SOA)} over $\Sigma$ is defined as a directed graph in which each node $v\in V \setminus \{src,snk\}$ is an SOIRE and all nodes are pairwise alphabet-disjoint SOIREs.

\section{Learning Algorithm}\label{sec3}
In this section, we give the learning algorithm $i$SOIRE, which efficiently infers an SORE from a set of positive samples $S$.
We show the major technical details of our algorithm in this section.
The input and output of the algorithm $i$SOIRE is a set of given samples and an SOIRE respectively.
The algorithm $i$SOIRE consists of two steps, constructing an SOA from samples, and converting the SOA into an SOIRE. Constructing an SOA from samples is introduced in Section~\ref{step1}.
Converting the SOA into an SOIRE is given in Section~\ref{step2}.

\begin{algorithm}[!htp]
  \caption{$i$SOIRE}\label{al1}
	\LinesNumbered
    \KwIn{a set of positive sample $S$}
    \KwOut{an SOIRE}
    Construct SOA $\mathcal{A}$ for $S$ using method 2T-INF~\cite{Garcia2002Inference};\\
    \textbf{return} Soa2Soire($S$, $\mathcal{A}$)

\end{algorithm}

\subsection{Constructing an SOA from Samples}\label{step1}
We use method 2T-INF~\cite{Garcia2002Inference} to construct SOA $\mathcal{A}$ for $S$. The algorithm 2T-INF~\cite{Garcia2002Inference} used in the algorithm is proved to construct a minimal-inclusion generalization of $S$. Here minimal-inclusion means that there is no other $SOA$ $\mathcal{A}$ such that $S \subseteq L(\mathcal{A})\subset L(SOA(S))$.

\begin{figure}
\center
\scalebox{1.0}{
\begin{tikzpicture}[->,>=stealth',shorten >=1pt,auto,node distance=0.7cm,
  thick,main node/.style={circle,fill=black!20,draw,minimum size=18pt}
]
  \node[main node,inner sep=1pt,minimum size=0pt,fill={rgb:red,1;green,2;blue,5}] (src) {\textcolor{white}{\textbf{$src$}}}; %,initial

   \node[main node] (b) [ right=of src] {$b$};
   \node[main node] (a) [above =of b] {$a$};

   %\node[state,thick,red] (e) [right =of b] {$e$};
   \node[main node] (e) [right =of b] {$e$};
   \node[main node] (n) [right =of e] {$n$};
   %\node[state] (n) [right =of e] {$n$};
   \node[main node] (g) [right =of n] {$g$};
\node[main node] (l) [below=of n] {$l$};
   %\node[state] (d) [above =of f] {$d$};
   \node[main node] (h) [below =of l] {$h$};

   \node[main node] (j) [above =of g] {$j$};
   \node[main node] (f) [above left =of j] {$f$};
   \node[main node] (c) [above =of f] {$c$};
   \node[main node] (m) [right =of f] {$m$};
   \node[main node] (d) [above =of c] {$d$};

   \node[main node] (k) [below =of g] {$k$};

   \node[main node,accepting,inner sep=1pt,minimum size=0pt,fill=black!30!green](snk) [right=of j] {\textcolor{white}{$snk$}};

    \path[->]
    %src=[a, b]
    (src) edge   (a)
    (src)  edge  (b)
    %a=[a, b, d, f]
    (a) edge  (b)
    (a) edge  (d)
    (a) edge  (f)
	   edge [loop above]  ()
    %b=[e, f, h]
    (b) edge  (e)
    (b) edge  (f)
    (b) edge  (h)
    %d=[snk, f, m]
    (d) edge [bend left] node [swap] {} (snk)
    (d) edge  [bend right] node [swap] {} (f)
    (d) edge  (m)
    %f=[c, d, snk, m]
    (f) edge  (snk)
    (f) edge  (c)
    (f) edge  [bend left] node [swap] {} (d)
    (f) edge  (m)
     %c=[d, m]
    (c) edge  (d)
    (c) edge  (m)
    %m=[c, d, snk]
    (m) edge  (snk)
    (m) edge  (c)
    (m) edge  (d)
    %e=[g, h, l, n]
    (e) edge [bend left] node [swap] {} (g)
    (e) edge  (h)
    (e) edge  (l)
    (e) edge  (n)
    %g=[snk, h, j, k]
    (g) edge  (snk)
    (g) edge  (h)
    (g) edge  (j)
    (g) edge  (k)
    %h=[e, g, k]
    (h) edge  (e)
    (h) edge  (g)
    (h) edge   (k)
    %j=[snk, j]
    (j) edge  (snk)
        edge [loop left]  ()
    %k=[snk]
    (k) edge  (snk)
    %l=[h]
    (l) edge  (h)
    %n=[g]
    (n) edge  (g);
\end{tikzpicture}
}
\caption{Constructing SOA $\bm{\mathcal{A}}$ for $\bm{S}$.}\label{SOA OF S}
\end{figure}
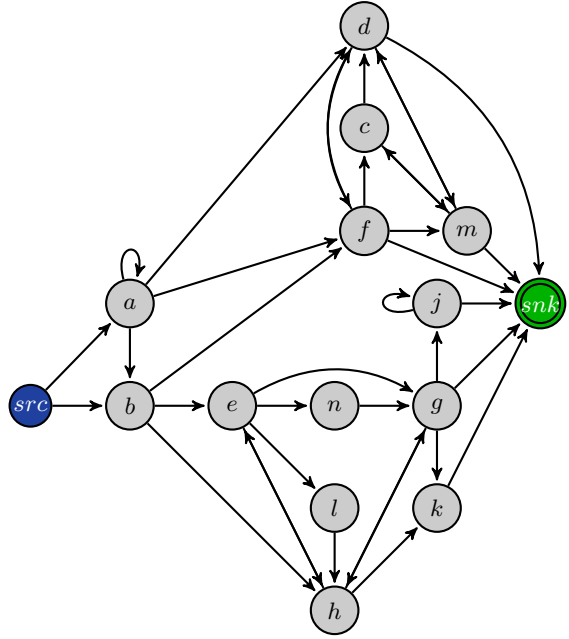

Here we give an example to show the execution process. Let $S$=$\{begk, aabengk,\\ abegjj, beg, hk, behgj, belhg, bheg, bfcmd, bfdm, afmcd, adf\}$. Using method \textit{2T-INF}, we construct the graph \textit{SOA(S)} shown in Figure~\ref{SOA OF S}.

\subsection{Converting the SOA into an SOIRE}\label{step2}
We use dot-notation to denote the application of subroutines. For a given SOA $\mathcal{A}$, we let $\mathcal{A}$.$src$ and $\mathcal{A}$.$snk$ denote the source and the sink of $\mathcal{A}$, respectively. We let $V$ be the set of vertices and $E$ the set of edges in $\mathcal{A}$, respectively.

\begin{itemize}
\item For any vertex $v \in V$, we let $\mathcal{A}$.pred($v$) denote the set of all predecessors of $v$ in $\mathcal{A}$; similarly, $\mathcal{A}$.succ($v$) denotes the set of all successors of $v$ in $\mathcal{A}$.
\item For any vertex $v \in V$, we let $\mathcal{A}$.reach($v$)  be the set of all vertices reachable from $v$.
\item ``first" returns all vertices $v$ such that the only predecessor of $v$ is the source in $\mathcal{A}$.
\item ``contract" on SOA $\mathcal{A}$ takes a subset $U$ of vertices of $\mathcal{A}$ and a label $\delta$. The procedure modifies $\mathcal{A}$ such that all vertices of $U$ are contracted to a single vertex and labeled $\delta$ (edges are moved accordingly).
\item ``extract" on SOA $\mathcal{A}$ takes as argument a set of vertices $U$ of $\mathcal{A}$; it does not modify $\mathcal{A}$, but returns a new SOA with copies of all vertices of $U$ as well as two new vertices for source and sink; all edges between vertices of $U$ are copied, all vertices in $U$ having an incoming
    edge in $\mathcal{A}$ from outside of $U$ have now an incoming edge from the new source, and all vertices in $U$ having
an outgoing edge in $\mathcal{A}$ to outside of $U$ have now an outgoing edge to the new sink.
\item ``addEpsilon" on SOA $\mathcal{A}$ adds a new vertex labeled $\varepsilon$; all
outgoing edges from the source to vertices that have
more than one predecessor (vertices, that are not in
the first-set) are redirected via this new vertex.
\item ``exclusive" on SOA $\mathcal{A}$ on argument $v$ (a vertex of $\mathcal{A}$) returns the set of all vertices $u$ such that, on any path
from the source to the sink that visits $u$, $v$ is necessarily visited previously. Intuitively, the exclusive set of a vertex $v$ is the set of all vertices exclusively reachable
from $v$, not from any other vertex incomparable to $v$.
\end{itemize}

Furthermore, we use the following eight subroutines or algorithms.
\begin{itemize}
\item ``plus" on label $\delta$ returns $\delta^+$.
\item ``or" on labels $\delta$ and $\delta'$ returns $\delta|\delta'$.
\item ``concatenate" on labels $\delta$ and $\delta'$ returns $\delta\cdot\delta'$.
\item ``filter" on a subset $U$ of vertices and a set of given sample $S$ returns a new subset $S'$.
        For string $s \in S$ each symbol of which is computed as follows: $\pi_s(U,s_i)=s_i$ if $s_i\in U$; $\pi_s(U,s_i)=\varepsilon$ otherwise. And the result is reduced by $x\varepsilon=\varepsilon x=x$. For example, let $U$=$\{b,c,r\}$ and $S=\{abgr,ebbdfc\}$, $S'= filter(U,S)=\{br,bbc\}$.
\item ``Merge" on a set of positive samples $S$ returns an expression $\zeta$ with interleaving.
\item   For a set of positive sample $S$, we let por($S$) denote the set of all partial order relations of each string in $S$ and cs($S$) denote the constraint set. The cs($S$) is defined as follows. $cs(S)=\{\langle x,y \rangle|\langle x,y \rangle \in por(S) ~and~ \langle y,x \rangle \in por(S)\}$.
\item ``combine" on a subset $U$ of vertices returns a new vertice, which combines all vertices in $U$ with interleaving operator.
        For example, let $U = \{a^*,b^+\}$, combine($U$) is $a^*\&b^+$.
\item ``$\rm{clique\_removal}$" on an undirected graph $G$ returns a maximum independent set (MIS). Finding an MIS of a graph $G$ is a NP-hard problem. Hence we use the method $\rm{clique\_removal}$() \cite{Boppana1992Approximating} to find an approximate result.
\end{itemize}

\begin{algorithm}[h]
  \caption{Soa2Soire}\label{alfinal}
	\LinesNumbered
    \KwIn{a set of positive sample $S$; an SOA $\mathcal{A}$ = ($V$,$E$)}
    \KwOut{an SOIRE}

	%$\mathscr{A}$ $\leftarrow$ $Soa(S)$\\
    \lIf {$|E|=0$ }{
        \textbf{return} $\varnothing$;\\
        }
    \lElseIf {$|V|=2$ }{
        \textbf{return} $\varepsilon$;\\
    }\ElseIf {$\mathcal{A}$ has a cycle}{
        Let $U$ be a strongly connected component of $\mathcal{A}$;\\
        \If {$|U|=1$ }{
        %$B_0$ $\leftarrow$ $\mathcal{A}$.extract($U$).bend()\\
        Let $v$ be the only vertice of $U$;\\
        $\mathcal{A}$.contract($U$,plus($v$.label()));\\
        }\lElse{
        %$S'$ $\leftarrow$ filter($U$, $S$)\\
        $\mathcal{A}$.contract($U$,Merge(filter($U$, $S$)));
        }
    }\ElseIf {$\mathcal{A}.\mathrm{succ(}\mathcal{A}.\mathrm{src)} \neq \mathcal{A}.\mathrm{first()}$}{
    $\mathcal{A}$.addEpsilon();
    }\ElseIf {$|\mathcal{A}.\mathrm{first()}|=1$ }{
    Let $v$ be the only successor of $src$;\\
    $\delta$ $\leftarrow$ $v$.label();\\
    $\mathcal{A}$.contract(\{$\mathcal{A}$.src,$v$\},src);\\
    $\delta'$ $\leftarrow$ Soa2Soire($S$,$\mathcal{A}$);\\
    \textbf{return} concatenate($\delta$,$\delta'$);
    }\ElseIf {$\exists v \in \mathcal{A}.\mathrm{first()}$, $\mathcal{A}.\mathrm{exclusive(}v\mathrm{)}$ $\neq$ \{$v$\}}{
        Let $v$ be such that $\mathcal{A}.\mathrm{exclusive(}v\mathrm{)}$ $\neq$ \{$v$\};\\
        $U$ $\leftarrow$ $\mathcal{A}.\mathrm{exclusive(}v\mathrm{)}$;\\
        $\mathcal{A}$.contract($U$,Soa2Soire($S$,$\mathcal{A}$.extract($U$)));
    }\Else{
    Let $u$,$v$ $\in$ $\mathcal{A}$.first() with $u$ $\neq$ $v$ s.t. $\mathcal{A}$.reach($u$) $\cap$ $\mathcal{A}$.reach($v$) is $\subseteq$-maximal;\\
    $\mathcal{A}$.contract(\{$u$,$v$\},or($u$.label(),$v$.label()));
    }

     \textbf{return} Soa2Soire($S$,$\mathcal{A}$);
\end{algorithm}
The algorithm Soa2Soire is given in Algorithm~\ref{alfinal}. The main procedures are as follows.
\begin{enumerate}
\item We first deal with all strongly connected looped components, replace each with a new vertex.
\item After the SOA is a directed acyclic graph (DAG), focus on the set $F$ of all vertices which can be reached from
the source directly, but not via other vertices; make sure that there are no vertices which can be reached directly and via other vertices (if necessary, add an
auxiliary node labeled $\varepsilon$).
\item Recurse on the sets of vertices exclusively reachable from a vertex in $F$ and contract these sets to vertices labeled with the result of the recursion.
\item Combine vertices in $F$ with ``or", recurse again on what
is exclusively reachable from this new vertex.
\item Once only one item is left in $F$, split it off and recurse on the remainder.
\end{enumerate}

Note that the algorithm introduces ``$?$" by way of constructing ``or $\varepsilon$". This can be cleaned up by postprocessing the
resulting SOIRE.

The algorithm Merge is given in Algorithm~\ref{al5}. The main procedures are as follows.
\begin{enumerate}
\item The first step (line 1): We first compute the constraint set $constraint\_tr$ using the function cs($S$).
\item The second step (line 4): We construct an undirected graph $G$ using element in $constraint\_tr$ as edges.
\item The third step (lines 5-8): We select a maximum independent set (MIS) of $G$, add it to list $all\_mis$ and delete the MIS and their related edges from $G$.
      The process is repeated until there exists no nodes in $G$.
\item The fourth step (lines 9-13): We get the sample set $S'$ using the function filter($mis$, $S$) for each MIS, and construct SOAs for sample sets by calling the
        algorithm 2T-INF~\cite{Garcia2002Inference}. Then convert SOAs into SOIREs using algorithm Soa2Soire.
\item The last step (line 14): We call the function combine to generate an expression $\zeta$ with interleaving operator.
\end{enumerate}

\begin{algorithm}[h]
  \caption{Merge}\label{al5}
	\LinesNumbered
    \KwIn{a set of positive sample $S$}
    \KwOut{an epression $\zeta$}
	$constraint\_tr$ $\leftarrow$ cs($S$);\\
     $U$ $\leftarrow \varnothing$;\\
    %$consist\_tr$ $\leftarrow$ CS($\overline{S}'$); $constraint\_tr$ $\leftarrow$ NCS($\overline{S}'$);\\
	$G$ $\leftarrow$ Graph($constraint\_tr$);\\
	$all\_mis$ $\leftarrow \varnothing$;\\
	\While {$|G.nodes()|> 0$ }{
    $W$ $\leftarrow$ $\rm{clique\_removal}$($G$)~\cite{Boppana1992Approximating};\\
    $G$ $\leftarrow$ $G \setminus W$;\\
     $all\_mis.append(G)$\\
    %\rm all $t^\prime$ in $\rm out_\sigma (s) \setminus \{ t\}$
	}

   % result $\leftarrow$ $\varepsilon$;\\

    \ForEach{$mis$ $\in$ $all\_mis$}{
    $S'$ $\leftarrow$ filter($mis$, $S$)\\
    Construct SOA $\mathcal{A}$ for $S'$ using method 2T-INF~\cite{Garcia2002Inference};\\
    $\delta$ $\leftarrow$ Soa2Soire($S'$,$\mathcal{A}$)\\  %$sub\_re$ $\leftarrow$ SOA2RE($\mathcal{A}_1$, $S''$) \\
    $U$.append($\delta$)\\
    }

  \textbf{return} $\zeta$ $\leftarrow$ combine($U$)\\
\end{algorithm}
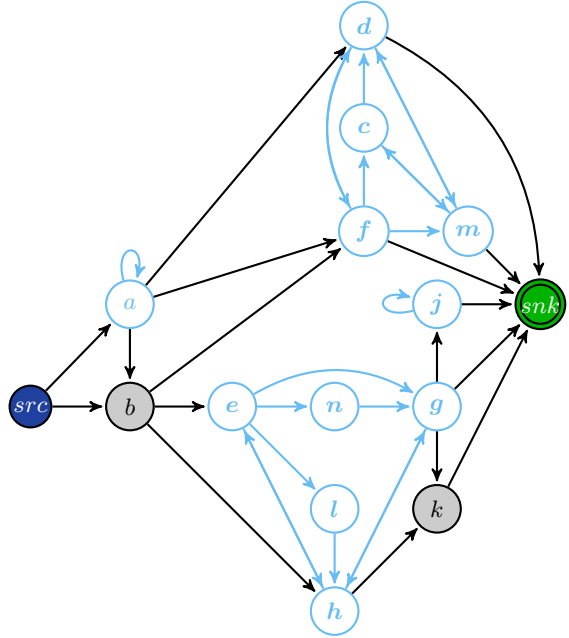
\begin{figure}[htp]
\center
\scalebox{1.0}{
\begin{tikzpicture}[->,>=stealth',shorten >=1pt,auto,node distance=0.7cm,
  thick,main node/.style={circle,fill=black!20,draw,minimum size=18pt}
]
  \node[main node,inner sep=1pt,minimum size=0pt,fill={rgb:red,1;green,2;blue,5}] (src) {\textcolor{white}{\textbf{$src$}}}; %,initial

   \node[main node] (b) [ right=of src] {$b$};
   \node[state,liyt_blue,minimum size=18pt] (a) [above =of b] {$a$};

   %\node[state,thick,red] (e) [right =of b] {$e$};
   \node[state,liyt_blue,minimum size=18pt] (e) [right =of b] {$\bm{e}$};
   \node[state,liyt_blue,minimum size=18pt] (n) [right =of e] {$\bm{n}$};
   %\node[state] (n) [right =of e] {$n$};
   \node[state,liyt_blue,minimum size=18pt] (g) [right =of n] {$\bm{g}$};
\node[state,liyt_blue,minimum size=18pt] (l) [below=of n] {$\bm{l}$};
   %\node[state] (d) [above =of f] {$d$};
   \node[state,liyt_blue,minimum size=18pt] (h) [below =of l] {$\bm{h}$};

   \node[state,liyt_blue,minimum size=18pt] (j) [above =of g] {$\bm{j}$};
   \node[state,liyt_blue,minimum size=18pt] (f) [above left =of j] {$\bm{f}$};
   \node[state,liyt_blue,minimum size=18pt] (c) [above =of f] {$\bm{c}$};
   \node[state,liyt_blue,minimum size=18pt] (m) [right =of f] {$\bm{m}$};
   \node[state,liyt_blue,minimum size=18pt] (d) [above =of c] {$\bm{d}$};

   \node[main node] (k) [below =of g] {$k$};

   \node[main node,accepting,inner sep=1pt,minimum size=0pt,fill=black!30!green](snk) [right=of j] {\textcolor{white}{\textbf{$snk$}}};

    \path[->]
    %src=[a, b]
    (src) edge   (a)
    (src)  edge  (b)
    %a=[a, b, d, f]
    (a) edge  (b)
    (a) edge  (d)
    (a) edge  (f)
	   edge [liyt_blue,loop above]  ()
    %b=[e, f, h]
    (b) edge  (e)
    (b) edge  (f)
    (b) edge  (h)
    %d=[snk, f, m]
    (d) edge [bend left] node [swap] {} (snk)
    (d) edge  [liyt_blue,bend right] node [swap] {} (f)
    (d) edge[liyt_blue]  (m)
    %f=[c, d, snk, m]
    (f) edge  (snk)
    (f) edge[liyt_blue]  (c)
    (f) edge  [liyt_blue,bend left] node [swap] {} (d)
    (f) edge[liyt_blue]  (m)
     %c=[d, m]
    (c) edge[liyt_blue]  (d)
    (c) edge[liyt_blue]  (m)
    %m=[c, d, snk]
    (m) edge  (snk)
    (m) edge[liyt_blue]  (c)
    (m) edge[liyt_blue]  (d)
    %e=[g, h, l, n]
    (e) edge [liyt_blue,bend left] node [swap] {} (g)
    (e) edge[liyt_blue]  (h)
    (e) edge[liyt_blue]  (l)
    (e) edge[liyt_blue]  (n)
    %g=[snk, h, j, k]
    (g) edge  (snk)
    (g) edge[liyt_blue]  (h)
    (g) edge  (j)
    (g) edge  (k)
    %h=[e, g, k]
    (h) edge[liyt_blue]  (e)
    (h) edge[liyt_blue]  (g)
    (h) edge   (k)
    %j=[snk, j]
    (j) edge  (snk)
        edge[liyt_blue,loop left]  ()
    %k=[snk]
    (k) edge  (snk)
    %l=[h]
    (l) edge[liyt_blue]  (h)
    %n=[g]
    (n) edge[liyt_blue]  (g);
\end{tikzpicture}
}
\caption{Four SCCs of SOA.}\label{SCC}
\end{figure}

Following the example in section \ref{step1}, there are four strongly connected components $U_1 = \{a\}$, $U_2 = \{j\}$, $U_3 = \{f,d,m,c\}$ and $U_4 = \{l,g,h,e,n\}$ shown in Figure~\ref{SCC}.
For strongly connected component (SCC) $U_1 = \{a\}$, because $|U_1|=1$, we use $\mathcal{A}$.contract($U_1$,plus($j$)) to modify $\mathcal{A}$ such that vertice $a$ is contracted to a new vertex $a^+$ and the self-loop is removed.
Similarly, we use $\mathcal{A}$.contract($U_2$,plus($j$)) to modify $\mathcal{A}$ such that vertice $j$ is contracted to a new vertex $j^+$ and the self-loop is removed (Figure~\ref{SCC1}).
For SCC $U_3$, because $|U_3|>1$, so we should call $\mathcal{A}$.contract($U_3$,Merge(filter($U_3$, $S$))). In this sub-process, we first compute the new sample set $S_1$=$\{fmcd,fcmd,df,fdm\}$ using function filter($U_3$,$S$).
Then we get cs($S_1$) = $\{\langle f,d\rangle, \langle d,f\rangle, \langle m,d\rangle, \langle d,m\rangle, \langle m,c\rangle, \langle c,m\rangle\}$ in the algorithm Merge. Next, we constructing undirected graph $G_1$ based on cs($S_1$) shown in Figure~\ref{ugcs1}.
We compute the set of all maximum independent sets (\textit{all\_mis} = $\{\{f,m\},\{c,d\}\}$) for Figure~\ref{ugcs1}.
We construct two SOAs using filter($\{f,m\}$, $S_1$) and filter($\{c,d\}$, $S_1$), respectively.
They are shown in Figure~\ref{soaaaa1} and Figure~\ref{soaaaa2}. We convert two SOAs into $fm^?$ and $c^?d$, respectively.
Then we get the new label $\zeta = fm^?\&c^?d$ using combine($fm^?$,$c^?d$).
We use $\mathcal{A}$.contract($U_3$,$\zeta$) to modify $\mathcal{A}$ such that all vertices of $U_3$ are contracted to a single vertex and labeled $\zeta$ (edges are moved accordingly) shown in Figure~\ref{SCC2}.
Similarly, we also call $\mathcal{A}$.contract($U$,Merge(filter($U_4$, $S$))). We first compute the new sample set $S_2=\{egh,eng,eg,\\elhg,ehg,heg\}$ using filter($U_4$,$S$).
Then we get cs($S_2$) = $\{\langle g,h\rangle, \langle h,g\rangle, \langle h,e\rangle, \langle e,h\rangle\}$ in the algorithm Merge. Next, we constructing undirected graph $G_2$ based on cs($S_2$) shown in Figure~\ref{ugcs2}.
We compute the set of all maximum independent sets $\{\{l,g,e,n\},\{h\}\}$ for Figure~\ref{ugcs2}.
We construct two SOAs using filter($\{l,g,e,n\}$, $S_2$) and filter($\{h\}$, $S_2$), respectively.
They are shown in Figure~\ref{soaaaa3} and Figure~\ref{soaaaa4}.
We convert two SOAs into $e(n|l)^?g$ and $h^?$, respectively.
Then we get the new label $\delta = e(n|l)^?g\&h^?$ using combine($e(n|l)^?g$,$h^?$).
We use $\mathcal{A}$.contract($U_4$,$\delta$) to modify $\mathcal{A}$ such that all vertices of $U_4$ are contracted to a single vertex and labeled $\delta$ (edges are moved accordingly) shown in Figure~\ref{SCC3}.
Continue to execute the remaining processes of the algorithm $i$SOIRE and we get the final inferred result $r$=$a^*b^?(fm^?\&c^?d|e(n|l)^?g\&h^?)(j^+|k)^?$.
\begin{figure}[htp]
\center
\scalebox{1.0}{
\begin{tikzpicture}[->,>=stealth',shorten >=1pt,auto,node distance=0.7cm,
  thick,main node/.style={circle,fill=black!20,draw,minimum size=18pt}
]
  \node[main node,inner sep=1pt,minimum size=0pt,fill={rgb:red,1;green,2;blue,5}] (src) {\textcolor{white}{\textbf{$src$}}}; %,initial

   \node[main node] (b) [ right=of src] {$b$};
      \node[main node,inner sep=1pt,minimum size=1pt] (a) [above =of b] {$\bm{a^+}$};

   %\node[state,thick,red] (e) [right =of b] {$e$};
   \node[state,liyt_blue,minimum size=18pt] (e) [right =of b] {$\bm{e}$};
   \node[state,liyt_blue,minimum size=18pt] (n) [right =of e] {$\bm{n}$};
   %\node[state] (n) [right =of e] {$n$};
   \node[state,liyt_blue,minimum size=18pt] (g) [right =of n] {$\bm{g}$};
\node[state,liyt_blue,minimum size=18pt] (l) [below=of n] {$\bm{l}$};
   %\node[state] (d) [above =of f] {$d$};
   \node[state,liyt_blue,minimum size=18pt] (h) [below =of l] {$\bm{h}$};

   \node[main node,inner sep=1pt,minimum size=1pt] (j) [above =of g] {$\bm{j^+}$};
   \node[state,liyt_blue,minimum size=18pt] (f) [above left =of j] {$\bm{f}$};
   \node[state,liyt_blue,minimum size=18pt] (c) [above =of f] {$\bm{c}$};
   \node[state,liyt_blue,minimum size=18pt] (m) [right =of f] {$\bm{m}$};
   \node[state,liyt_blue,minimum size=18pt] (d) [above =of c] {$\bm{d}$};

   \node[main node] (k) [below =of g] {$k$};

   \node[main node,accepting,inner sep=1pt,minimum size=0pt,fill=black!30!green](snk) [right=of j] {\textcolor{white}{\textbf{$snk$}}};

    \path[->]
    %src=[a, b]
    (src) edge   (a)
    (src)  edge  (b)
    %a=[a, b, d, f]
    (a) edge  (b)
    (a) edge  (d)
    (a) edge  (f)
    %b=[e, f, h]
    (b) edge  (e)
    (b) edge  (f)
    (b) edge  (h)
    %d=[snk, f, m]
    (d) edge [bend left] node [swap] {} (snk)
    (d) edge  [liyt_blue,bend right] node [swap] {} (f)
    (d) edge[liyt_blue]  (m)
    %f=[c, d, snk, m]
    (f) edge  (snk)
    (f) edge[liyt_blue]  (c)
    (f) edge  [liyt_blue,bend left] node [swap] {} (d)
    (f) edge[liyt_blue]  (m)
     %c=[d, m]
    (c) edge[liyt_blue]  (d)
    (c) edge[liyt_blue]  (m)
    %m=[c, d, snk]
    (m) edge  (snk)
    (m) edge[liyt_blue]  (c)
    (m) edge[liyt_blue]  (d)
    %e=[g, h, l, n]
    (e) edge [liyt_blue,bend left] node [swap] {} (g)
    (e) edge[liyt_blue]  (h)
    (e) edge[liyt_blue]  (l)
    (e) edge[liyt_blue]  (n)
    %g=[snk, h, j, k]
    (g) edge  (snk)
    (g) edge[liyt_blue]  (h)
    (g) edge  (j)
    (g) edge  (k)
    %h=[e, g, k]
    (h) edge[liyt_blue]  (e)
    (h) edge[liyt_blue]  (g)
    (h) edge   (k)
    %j=[snk, j]
    (j) edge  (snk)
    %k=[snk]
    (k) edge  (snk)
    %l=[h]
    (l) edge[liyt_blue]  (h)
    %n=[g]
    (n) edge[liyt_blue]  (g);
\end{tikzpicture}
}
\caption{Dealing with SCC $\bm{U_1}$ and $\bm{U_2}$ of SOA $\bm{\mathcal{A}}$.}\label{SCC1}
\end{figure}
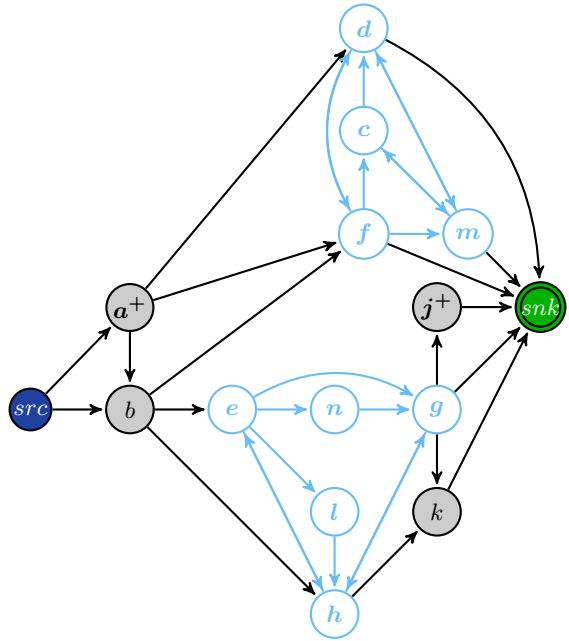
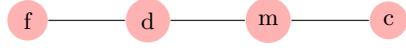
\begin{figure}[htp]
\center
\scalebox{1.0}{
\begin{tikzpicture}[> = stealth, % arrow head style
shorten > = 0.1pt, % don't touch arrow head to node
auto,
node distance = 0.6cm, % distance between nodes
%semithick, % line style
scale=.8,auto=left,every node/.style={circle,fill=red!30}]  %blue%
\node (n1) at (0,0)     {f};
\node (n2) at (2,0)    {d};
\node (n3) at (4,0)    {m};
\node (n4) at (6,0)    {c};
\draw (n1)--(n2);
\draw (n2)--(n3);
\draw (n3)--(n4);
\end{tikzpicture}
}
\caption{Constructing undirected graph $\bm{G_1}$.}
\label{ugcs1}
\end{figure}
\begin{figure}[htp]
\center
\scalebox{1.0}{
\begin{tikzpicture}[->,>=stealth',shorten >=1pt,auto,node distance=0.7cm,
  thick,main node/.style={circle,fill=black!20,draw,minimum size=18pt}
]

  \node[main node,inner sep=1pt,minimum size=0pt,fill={rgb:red,1;green,2;blue,5}] (src) {\textcolor{white}{\textbf{$src$}}}; %,initial
   \node[main node] (f) [right =of src] {$f$};
\node[main node] (m) [right =of f] {$m$};
   \node[main node,accepting,inner sep=1pt,minimum size=0pt,fill=black!30!green](snk) [right =of m] {\textcolor{white}{\textbf{$snk$}}};
\path[->]

    (src) edge   (f)
    (f) edge   (m)
    (f)  edge [bend right=45] node [swap] {} (snk)
    (m) edge   (snk);
\end{tikzpicture}
}
\caption{Constructing SOA $\bm{\mathcal{A}_1}$ of filter($\bm{\{f,m\}}$, $\bm{S_1}$).}
\label{soaaaa1}
\end{figure}
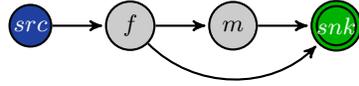

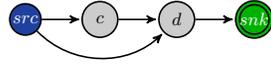
\begin{figure}[htp]
\center
\scalebox{0.75}{
\begin{tikzpicture}[->,>=stealth',shorten >=1pt,auto,node distance=0.7cm,
  thick,main node/.style={circle,fill=black!20,draw,minimum size=18pt}
]
  \node[main node,inner sep=1pt,minimum size=0pt,fill={rgb:red,1;green,2;blue,5}] (src) {\textcolor{white}{\textbf{$src$}}}; %,initial
   \node[main node] (c) [right =of src] {$c$};
\node[main node] (d) [right =of c] {$d$};
   \node[main node,accepting,inner sep=1pt,minimum size=0pt,fill=black!30!green](snk) [right =of d] {\textcolor{white}{\textbf{$snk$}}};
\path[->]

    (src) edge   (c)
    (c) edge   (d)
    (src)  edge [bend right=45] node [swap] {} (d)
    (d) edge   (snk);
\end{tikzpicture}
}
\caption{Constructing SOA $\bm{\mathcal{A}_2}$ of filter($\bm{\{c,d\}}$, $\bm{S_1}$).}
\label{soaaaa2}
\end{figure}

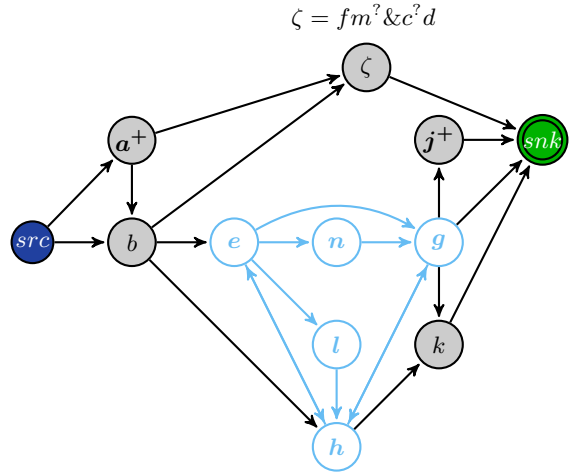
\begin{figure}[htp]
\center
\scalebox{1.0}{
\begin{tikzpicture}[->,>=stealth',shorten >=1pt,auto,node distance=0.7cm,
  thick,main node/.style={circle,fill=black!20,draw,minimum size=18pt}
]
  \node[main node,inner sep=1pt,minimum size=0pt,fill={rgb:red,1;green,2;blue,5}] (src) {\textcolor{white}{\textbf{$src$}}}; %,initial

   \node[main node] (b) [ right=of src] {$b$};
      \node[main node,inner sep=1pt,minimum size=1pt] (a) [above =of b] {$\bm{a^+}$};

   %\node[state,thick,red] (e) [right =of b] {$e$};
   \node[state,liyt_blue,minimum size=18pt] (e) [right =of b] {$\bm{e}$};
   \node[state,liyt_blue,minimum size=18pt] (n) [right =of e] {$\bm{n}$};
   %\node[state] (n) [right =of e] {$n$};
   \node[state,liyt_blue,minimum size=18pt] (g) [right =of n] {$\bm{g}$};
\node[state,liyt_blue,minimum size=18pt] (l) [below=of n] {$\bm{l}$};
   %\node[state] (d) [above =of f] {$d$};
   \node[state,liyt_blue,minimum size=18pt] (h) [below =of l] {$\bm{h}$};

   \node[main node,inner sep=1pt,minimum size=1pt] (j) [above =of g] {$\bm{j^+}$};
   \node[main node,minimum size=18pt] (f) [above left =of j] {$\zeta$};
   \node[main node] (k) [below =of g] {$k$};

   \node[main node,accepting,inner sep=1pt,minimum size=0pt,fill=black!30!green](snk) [right=of j] {\textcolor{white}{\textbf{$snk$}}};
   \node[fill=white]  (pie) at (4.4,3.0) {$\zeta = fm^?\&c^?d$};%Inference Engine

    \path[->]
    %src=[a, b]
    (src) edge   (a)
    (src)  edge  (b)
    %a=[a, b, d, f]
    (a) edge  (b)
    (a) edge  (f)
    %b=[e, f, h]
    (b) edge  (e)
    (b) edge  (f)
    (b) edge  (h)
    %d=[snk, f, m]

    %f=[c, d, snk, m]
    (f) edge  (snk)
    (e) edge [liyt_blue,bend left] node [swap] {} (g)
    (e) edge[liyt_blue]  (h)
    (e) edge[liyt_blue]  (l)
    (e) edge[liyt_blue]  (n)
    %g=[snk, h, j, k]
    (g) edge  (snk)
    (g) edge[liyt_blue]  (h)
    (g) edge  (j)
    (g) edge  (k)
    %h=[e, g, k]
    (h) edge[liyt_blue]  (e)
    (h) edge[liyt_blue]  (g)
    (h) edge   (k)
    %j=[snk, j]
    (j) edge  (snk)
    %k=[snk]
    (k) edge  (snk)
    %l=[h]
    (l) edge[liyt_blue]  (h)
    %n=[g]
    (n) edge[liyt_blue]  (g);
\end{tikzpicture}
}
\caption{Dealing with SCC $\bm{U_3}$ of SOA $\bm{\mathcal{A}}$.}\label{SCC2}
\end{figure}

\begin{figure}[htp]
\center
\scalebox{1.0}{
\begin{tikzpicture}[> = stealth, % arrow head style
shorten > = 0.1pt, % don't touch arrow head to node
auto,
node distance = 0.6cm, % distance between nodes
%semithick, % line style
scale=.8,auto=left,every node/.style={circle,fill=red!30}]  %blue%
\node (n0) at (0,0)     {l};
\node (n1) at (2,0)     {g};
\node (n2) at (4,0)    {h};
\node (n3) at (6,0)    {e};
\node (n4) at (8,0)    {n};
\draw (n1)--(n2);
\draw (n2)--(n3);
%\draw [red,very thick](n4)--(n5);
%\draw [red,very thick] (n1) .. controls (-2,-2.5)  ..(n4);
%\draw [red,very thick](n1) .. controls (2,-3.5)     ..(n5);
\end{tikzpicture}
}
\caption{Constructing undirected graph $\bm{G_2}$.}
\label{ugcs2}
\end{figure}
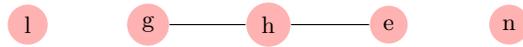

\begin{figure}[htp]
\center
\scalebox{1.0}{
\begin{tikzpicture}[->,>=stealth',shorten >=1pt,auto,node distance=0.7cm,
  thick,main node/.style={circle,fill=black!20,draw,minimum size=18pt}
]
   \node[main node,inner sep=1pt,minimum size=0pt,fill={rgb:red,1;green,2;blue,5}] (src) {\textcolor{white}{\textbf{$src$}}}; %,initial
   \node[main node] (e) [right =of src] {$e$};
   \node[main node] (l) [above right =of e] {$l$};
   \node[main node] (g) [below right =of l] {$g$};
   \node[main node] (n) [below right =of e] {$n$};
   \node[main node,accepting,inner sep=1pt,minimum size=0pt,fill=black!30!green](snk) [right =of g]  {\textcolor{white}{\textbf{$snk$}}};
\path[->]
    (src) edge   (e)
    (e) edge   (l)
    (e) edge   (g)
    (e) edge   (n)
    (l) edge   (g)
    (n) edge   (g)
    (g) edge   (snk);
\end{tikzpicture}
}
\caption{Constructing SOA $\bm{\mathcal{A}_3}$.}
\label{soaaaa3}
\end{figure}
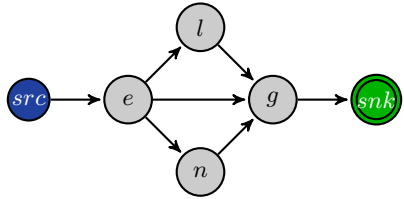

\begin{figure}[htp]
\center
\scalebox{1.0}{
\begin{tikzpicture}[->,>=stealth',shorten >=1pt,auto,node distance=0.7cm,
  thick,main node/.style={circle,fill=black!20,draw,minimum size=18pt}
]
     \node[main node,inner sep=1pt,minimum size=0pt,fill={rgb:red,1;green,2;blue,5}] (src) {\textcolor{white}{\textbf{$src$}}}; %,initial
   \node[main node] (h) [right =of src] {$h$};
   \node[main node,accepting,inner sep=1pt,minimum size=0pt,fill=black!30!green](snk) [right =of h] {\textcolor{white}{\textbf{$snk$}}};
\path[->]

    (src) edge   (h)
    (src)  edge [bend left=45] node [swap] {} (snk)
    %(src)  edge [bend right] node [swap] {} (snk)
    (h) edge   (snk);
\end{tikzpicture}
}
\caption{Constructing SOA $\bm{\mathcal{A}_4}$.}
\label{soaaaa4}
\end{figure}
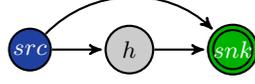
\begin{figure}[htp]
\center
\scalebox{1.0}{
\begin{tikzpicture}[->,>=stealth',shorten >=1pt,auto,node distance=0.7cm,
  thick,main node/.style={circle,fill=black!20,draw,minimum size=18pt}
]
  \node[main node,inner sep=1pt,minimum size=0pt,fill={rgb:red,1;green,2;blue,5}] (src) {\textcolor{white}{\textbf{$src$}}}; %,initial

   \node[main node] (b) [ right=of src] {$b$};
      \node[main node,inner sep=1pt,minimum size=1pt] (a) [above =of b] {$\bm{a^+}$};

   \node[main node] (e) [right =of b] {$\delta$};
   \node[main node,inner sep=1pt,minimum size=1pt] (j) [right =of e] {$\bm{j^+}$};
   \node[main node] (k) [right =of j] {$k$};

   \node[main node,accepting,inner sep=1pt,minimum size=0pt,fill=black!30!green](snk) [above=of j] {\textcolor{white}{\textbf{$snk$}}};
   \node[main node,minimum size=18pt] (f) [left =of snk] {$\zeta$};
   \node  (pie) at (2.8,2.0) {$\zeta = fm^?\&c^?d$};%Inference Engine
   \node  (pie) at (2.7,-0.7) {$\delta = e(n|l)^?g\&h^?$};%Inference Engine

    \path[->]
    %src=[a, b]
    (src) edge   (a)
    (src)  edge  (b)
    %a=[a, b, d, f]
    (a) edge  (b)
    (a) edge  (f)
    (e) edge  (j)
    (e) edge  (snk)
    (e) edge[bend right]  (k)
    %b=[e, f, h]
    (b) edge  (e)
    (b) edge  (f)
    %d=[snk, f, m]

    %f=[c, d, snk, m]
    (f) edge  (snk)
    %g=[snk, h, j, k]
    %h=[e, g, k]
    %j=[snk, j]
    (j) edge  (snk)
    %k=[snk]
    (k) edge  (snk);
\end{tikzpicture}
}
\caption{Dealing with SCC $\bm{U_4}$ of SOA $\bm{\mathcal{A}}$.}\label{SCC3}
\end{figure}
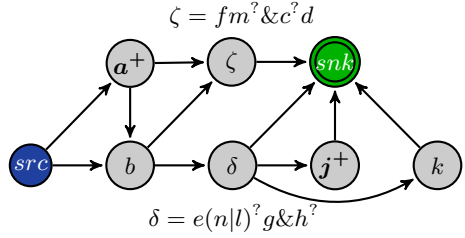

%strongly

\iffalse
\textbf{Algorithm Analysis} an SOA $\mathcal{A}$ = ($V$,$E$), let $n$=$|V|$ and $m$=$|E|$. It costs time $O(n)$ to find all nodes with loops and $O(m+n)$ to find all strongly connected components. The time complexity of $clique\_removal()$ is $O(n^2+m)$. For each NTSCC, computation of $all\_mis$ costs time $O(n^3+m)$. For each $mis$, there is no NTSCC at all. Hence \textit{Repair()} only costs time $O(n^3+m)$. The number of NTSCCs in a SOA is finite. Then computation of $all\_mis$ for all NTSCCs also costs time $O(n^3+m)$. Assigning level numbers and computing all skip levels will be finished in time $O(m+n)$. All nodes will be converted into specific chain factors in $O(n)$. Therefore, the time complexity of $GenESIRE$ is $O(n^3+m)$.
\fi

\section{Experiments}\label{sec4}

% 实验主要针对三个方面, 强调对比实验
In this section, we conduct a series of experiments to analyze the practicability of SOIRE, and compare algorithm $i$SOIRE with not only the learning algorithms from ongoing researches but also the industrial-level tools used in real world. In terms of \textit{preciseness} and \textit{conciseness}, our work has achieved satisfying results compared with existing methods, reaching higher preciseness with less description length.
Specifically, indicators
\textit{Language Size} (\bm{$|\mathcal{L}(r)|$})~\cite{DBLP:journals/tweb/BexGNV10} and \textit{datacost} (\textbf{DC})~\cite{DBLP:journals/tweb/BexGNV10} are used to measure preciseness, while \bm{$\mathcal{L}en$}~\cite{DBLP:conf/adma/LiMC18} and Nesting Depth (\textbf{ND})~\cite{DBLP:conf/apweb/LiZPC16} for conciseness. Similar as the discussion of $|\mathcal{L}(r)|$ and $\mathcal{L}en$ above, we have that larger the value of DC (ND) is, more precise (concise) the regular expression will be.
 \textit{Language Size}~\cite{DBLP:journals/tweb/BexGNV10}, denoted by $|\mathcal{L}(r)|$, is defined as:
 $$|\mathcal{L}(r)|=\sum_{\ell=1}^{\ell_{max}} |L^\ell(r)|,$$ where $|L^\ell(r)|$ is the size of subset containing words with length $\ell$ in $L(r)$. Generally, $L(r)$ is an infinite language with infinitely large value of $\ell$, it is of course impossible to take all words into account. Hence, we only consider the word length $\ell$ up to a maximum value: $\ell_{max}=2m+1$ where $m$ is the length of $r$ excluding $\varepsilon$, $\varnothing$ and regular expression operators. \textit{Language Size} ($|\mathcal{L}(r)|$) can well measure the preciseness of a regular expression. Smaller the value of $|\mathcal{L}(r)|$ is, more precise the regular expression will be.
\textit{datacost} (\textbf{DC})~\cite{DBLP:journals/tweb/BexGNV10}, is defined as:
$$datacost(r,S) = \sum_{\ell=1}^{\ell_{max}}\left( 2\times log_2 \ell + log_2 \binom {|L^\ell(r)|} {|S^\ell|}\right),$$
where $\ell_{max}=2m+1$ and $|L^\ell(r)|$ as before, $|S^\ell|$ is the number of words in $S$ that have length $\ell$.
Smaller the value of \textbf{DC} is, more precise the regular expression will be. $\mathcal{L}en$~\cite{DBLP:conf/adma/LiMC18} is defined as:
$$\mathcal{L}en=n\times \lceil \log_2 (|\Sigma|+|\mathcal{M}|)\rceil,$$%\ast
where $|\Sigma|$ is the number of distinct symbols occurring in regular expression $r$, $\mathcal{M}$ is the set of metacharacters $\{|,\cdot,\&,?,*,+,\\(,)\}$ and $n$ is the length of $r$ including symbols and metacharacters. An expression with a smaller value of $\mathcal{L}en$ is more concise.
Nesting Depth (\textbf{ND})~\cite{DBLP:conf/apweb/LiZPC16} is defined as:

\begin{itemize}
    \item ND($r$) = 0, if $r$ = $\varepsilon$, $\varnothing$ or $a$ for $a \in \Sigma$.
    \item  ND($r$) =  ND($r_1$) + 1, if $r = r_1^*$, $r = r_1^?$ or $r = r_1^+$, where $r_1$ is a regular expression over $\Sigma$.
    \item ND($r$) = $max\{$ND($r_1$),ND($r_2$)$\}$, if $r = r_1 | r_2$, $r = r_1 \cdot r_2$ or $r = r_1 \& r_2$, where $r_1$ and $r_2$ are regular expressions over $\Sigma$.
\end{itemize}
The learning algorithms compared in experiments are Soa2Sore~\cite{DBLP:journals/mst/FreydenbergerK15} and Soa2Chare~\cite{DBLP:journals/mst/FreydenbergerK15}, GenEchare~\cite{Feng20},
$learner_{DME}^+$ ~\cite{DBLP:journals/corr/CiucanuS13},
conMiner~\cite{DBLP:conf/apweb/PengC15}, GenICHARE~\cite{DBLP:conf/pakdd/ZhangLCDC18} and GenESIRE~\cite{DBLP:conf/er/LiZXMC18}. The industrial tools which are capable of supporting inference of XML schemas used in this section include IntelliJ IDEA\footnote{\url{https://www.jetbrains.com/idea/}}, Liquid Studio\footnote{\url{https://www.liquid-technologies.com/}}, Trang\footnote{\url{http://www.thaiopensource.com/relaxng/trang.html}}, and InstanceToSchema\footnote{\url{http://www.xmloperator.net/i2s/}}.

% the new subclass of regular expressions.
% In addition, we analyze the effectiveness of our inference algorithm.
% The dataset that our experiments base on consists of  schema files with 13,946 regular expressions.

{\color{bella} For the massive comparative experiments, we conduct the experiments based on two kinds of datasets: small dataset (i.e., \textit{mastersthesis}) and large dataset (i.e, \textit{www})} of XML documents, which are both extracted from DBLP. DBLP is a {\color{bella} data-centered database} of information on major computer science journals and proceedings. We download the file of version \textit{dblp-2015-03-02.xml.gz}\footnote{\url{http://dblp.org/xml/release/dblp-2015-03-02.xml.gz}}. \textit{mastersthesis} and \textit{www} are two elements chosen from DBLP with $ 5$ (small) and $2,000,226$ (large) samples, respectively.

All of our experiments are conducted on a machine with 16 cores Intel Xeon CPU E5620 @ 2.40GHz with 12M Cache, 24G RAM, OS: Windows 10.
\subsection{Usage of SOIRE in Practice}

\begin{figure}
\centering
\begin{tikzpicture}[scale = 0.7]
  \definecolor{myblue}{HTML}{1F7ED2}
  \begin{axis}[
    axis x line*=bottom,
    axis y line*=none,
    every outer y axis line/.append style={draw=none},
    every y tick/.append style={draw=none},
    ymin=0,
    ymax=100,
     ytick={0, 10, ..., 100},
    ymajorgrids,
    y grid style={densely dotted, line cap=round},
    ylabel={Percent in ALL REs(\%)},
    nodes near coords,
     xtick=data,
    xticklabel style={rotate=60,  xshift=0.0cm ,font=\scriptsize},
   % x tick label style={rotate=-90, xshift=1cm,yshift=-0.2cm, anchor=south,font=\small},%,align=center
   symbolic x coords={DME, CHARE, SORE, $k$-ORE, SIRE, ICRE, ICHARE, ESIRE, SOIRE},
   height=5cm,width=11cm,
  ]
    \addplot[
      ybar,
      draw=none,
      fill=myblue,
    ] coordinates {
      (DME, 50.62)
      (CHARE, 55.04)
      %(eSimplified CHARE, 56.10)%%
      (SORE, 58.30)
      ($k$-ORE, 76.11)
      (SIRE, 76.83)
      (ICRE, 85.55)
      (ICHARE, 85.78)
      (ESIRE, 87.56)
      (SOIRE, 93.24)
    };
  \end{axis}
\end{tikzpicture}

\caption{\textbf{The proportion of subclasses on Relax NG.} {The dataset used for this statistical experiment is acquired from~\cite{DBLP:conf/ideas/LiCMDC18}, with $509,267$ regular expressions from $4,526$ Rleax NG schemas.}}
    \label{fig:stats}
\end{figure}
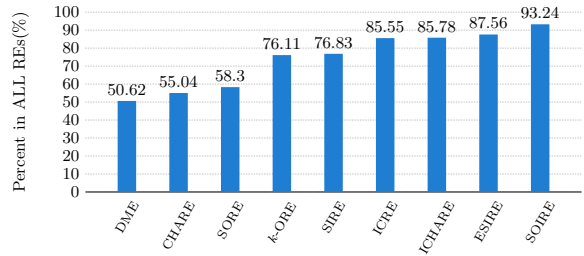

Though interleaving is indispensable in data-centric applications, the lack of research on it is still a concern.
% Here, we statistically demonstrate the practicality of our work.
In Figure~\ref{fig:stats}, we visualized the coverage rates of regular expressions covered by different subclasses on Relax NG. We can see that the initial subclass, DME, only covers 50.62\%. Then the proportions show an upward trend, reaching more than 85.55\% (ICRE, ICHARE, ESIRE). Compared with their coverage, SOIRE covers 93.24\%, which is 5.68\% more than the second largest proportion. Therefore, the experimental result reveals the high practicality of SOIRE, and its strong support for interleaving.

\subsection{Analysis of Inference Results}
% Second, based on different measures (\textit{ND}, \textit{Len} and language size), we compare the inferred results of iKOIRE with other methods based on our data sets.

To better illustrate the performance of our work, we first compare the inferred results of our work with that of existing learning algorithms and industrial tools in real world.
To save space, we use the short names of words and the list of abbreviations is shown in Table~\ref{abb1}.
The experimental results are shown in Table 2-5.%~\ref{tab:pos1111}, Table~\ref{tab:pos2222}, Table~\ref{tab:pos3333} and Table~\ref{tab:pos4444}.
\sisetup{round-mode=places
        ,round-precision=2
        ,scientific-notation=false
    }

\begin{table}[h]
\centering
\scriptsize{
\renewcommand\arraystretch{2.0}
 \setlength{\belowcaptionskip}{3pt}
\caption{The list of abbreviations for words in DBLP.}
\center
\begin{tabular}{|p{1.2cm}<{\centering}|c||p{1.2cm}<{\centering}|c||p{1.2cm}<{\centering}|c|}
% xyz
\hline%
\textbf{Word}&\textbf{Abbr.} & \textbf{Word}&\textbf{Abbr.} & \textbf{Word}&\textbf{Abbr.}  \\
\hline
  author&a&editor&b&title&c\\
  \hline
  booktitle&d&pages&e&year&f\\
  \hline
  address&g&journal&h&volume&i\\
   \hline
  number&j&month&k&url&l\\
 \hline
 ee&m&cdrom&n&cite&o\\
 \hline
 publisher&p&note&q&crossref&r\\
 \hline
 isbn&s&series&t&school&u\\
 \hline
 chapter&v&publnr&w&&\\
 \hline
\end{tabular}
\label{abb1} }
\end{table}

We can see from Table~\ref{tab:pos2222} that for dataset \textit{mastersthesis}, the first six algorithms/tools (Liquid Studio, Soa2Sore, Soa2Chare, GenEchare, IntelliJ IDEA and Trang) reach high conciseness at enormous cost of $|\mathcal{L}(r)|$, from unaffordable \num{1.5673E+10} to \num{1.6383E+4}.
 Algorithms/tools InstanceToSchema, $learner_{DME}^+$ and conMiner have highest conciseness, with 52 for $\mathcal{L}en$, yet their preciseness is not the highest among these algorithms.
  Finally, the last three algorithms including $i$SOIRE reach the performance at the same level, with highest preciseness and the equal magnitude of conciseness. From the table we can draw a conclusion that though interleaving could improve the preciseness, the former one sacrifices the conciseness to some degree.

  \begin{table}[htbp]
\sisetup{round-mode=places
        ,round-precision=2
        ,scientific-notation=false
    }
\center
\scriptsize
\renewcommand\arraystretch{2.0}
\caption{Expressions of inference using different learning algorithms/inference tools on mastersthesis.}

%\begin{tabular}{c|[1pt]p{5.6cm}<{\centering}}
\begin{tabular}{|p{2.2cm}<{\centering}|p{5.0cm}<{\centering}|}
\hline
\footnotesize{\textbf{Method}}&\footnotesize{\textbf{Regular Expression}}
\\\hline
Liquid Studio
 &$\rm{(a|c|f|u|l|m)^+}$
\\
Soa2Sore
&$\rm{acfu(l|m)^*}$
\\
Soa2Chare
&$\rm{acfu(l|m)^*}$
\\
GenEchare
&$\rm{acfu(l|m)^*}$
\\
IntelliJ IDEA
&$\rm{acfu(l|m)^*}$
\\
Trang
&$\rm{acfu(l|m)^*}$
\\
InstanceToSchema
&$\rm{a\&c\&f\&l^?\&m^?\&u}$
\\
$learner_{DME}^+$
&$\rm{a\&c\&f\&l^?\&m^?\&u}$
\\
conMiner
&$\rm{acful^?\&m^?}$
\\
GenICHARE
&	$\rm{acfu(l^?\&m^?)}$
\\
GenESIRE
&	$\rm{acfu(l^?\&m^?)}$
\\
$i\mathrm{SOIRE}$
&	$\rm{acfu(l^?\&m^?)}$
\\
\hline
\end{tabular}
\label{tab:pos1111}
\end{table}

\begin{table}[htbp]
\sisetup{round-mode=places
        ,round-precision=2
        ,scientific-notation=false
    }
\center
\scriptsize
\renewcommand\arraystretch{2.0}
\caption{Results of inference using different learning algorithms/inference tools on mastersthesis.}

%\begin{tabular}{c|[1pt]p{5.6cm}<{\centering}}
\begin{tabular}{|p{2.2cm}<{\centering}|p{1.6cm}<{\centering}|p{1.2cm}<{\centering}|p{0.6cm}<{\centering}|p{0.6cm}<{\centering}|}
\hline
\footnotesize{\textbf{Method}}&\footnotesize{\bm{$|\mathcal{L}(r)|$}}
&\footnotesize{\textbf{DC}}&\footnotesize{\bm{$\mathcal{L}en$}}&\footnotesize{\textbf{ND}}
\\\hline
Liquid Studio
 &\num{1.5673E+10}&$122.880$&$56$	& $1$
\\
Soa2Sore
&\num{1.6383E+4}&$67.657$&	$56$	& $1$
\\
Soa2Chare
&\num{1.6383E+4}&$67.657$&	$56$	& $1$
\\
GenEchare
&\num{1.6383E+4}&$67.657$&	$56$	& $1$
\\
IntelliJ IDEA
&\num{1.6383E+4}&$67.657$&	$56$	& $1$
\\
Trang
&\num{1.6383E+4}&$67.657$&	$56$	& $1$
\\
InstanceToSchema
&$984$&	$102.446$&	$52$&$1$
\\
$learner_{DME}^+$
&$984$&	$102.446$&	$52$&$1$
\\
conMiner
&$13$	&$72.886$&	$52$	&$1$
\\
$\bm{\mathrm{GenICHARE}}$
&	$\bm{5}$	&$\bm{65.072}$&	$\bm{60}$& $\bm{1}$
\\
$\bm{\mathrm{GenESIRE}}$
&	$\bm{5}$	&$\bm{65.072}$&	$\bm{60}$& $\bm{1}$
\\
$\bm{i\mathrm{SOIRE}}$
&	$\bm{5}$	&$\bm{65.072}$&	$\bm{60}$& $\bm{1}$
\\
\hline
\end{tabular}

\label{tab:pos2222}
\end{table}

For the second dataset (Table~\ref{tab:pos4444}), the advantage of our work is more outstanding. Without supporting the usage of interleaving, the previous eleven algorithms/tools have huge $|\mathcal{L}(r)|$ and DC, from \num{1.1111E+21} to \num{4.3899E+11} and from $15158.773$ to $8479.873$, respectively. Among them, Liquid Studio, Soa2Chare and IntelliJ IDEA have the shortest $\mathcal{L}en$, which are 120, while $learner_{DME}^+$ and ESIRE have the longest, which are 175. Soa2Sore has the deepest ND~\cite{DBLP:conf/apweb/LiZPC16}, with 3,
followed by Liquid Studio, GenEchare, GenICHARE and GenESIRE, with 2 nestings. On the other hand, the algorithms/tools which support interleaving have smaller values on average. Especially for the indicator $|\mathcal{L}(r)|$, the magnitudes are much smaller than that of the first group of methods. It is noteworthy that our work reaches almost the same conciseness with much less values of $|\mathcal{L}(r)|$(\num{1.83783E+11}) and DC($7599.996$).

\iffalse
For the second dataset (Table~\ref{tab:pos4444}), the advantage of our work is more outstanding. While other algorithms/tools have tremendous $|\mathcal{L}(r)|$ and DC, our work only need \num{1.83783E+11} for $|\mathcal{L}(r)|$ and $7599.996$ for DC. It means that no matter how complicated the regular expression is, our work can still achieve stable results with both high preciseness. Though the $\mathcal{L}en$ of our work is the largest, the gap is relatively small, especially compared with the huge gap in $|\mathcal{L}(r)|$.

\fi

\begin{table}[htbp]
\center
\scriptsize
\renewcommand\arraystretch{2.0}
\caption{Expressions of inference using different learning algorithms/inference tools on www.}
%\begin{tabular}{c|[1pt]p{5.6cm}<{\centering}}
\begin{tabular}{|p{2.2cm}<{\centering}|p{5.0cm}<{\centering}|}
\hline
\footnotesize{\textbf{Method}}&\footnotesize{\textbf{Regular Expression}}
\\\hline
Liquid Studio
&$\rm{(a|c|l^+|q^+|o|b|f|m|d|r)^+}$
\\
Soa2Sore
&$\rm{b^*(a^*(c(m^?|d))^?(l|q|f|o)^*)^+|r}$
\\
Soa2Chare
&$\rm{b^*r^?(m|o|f|a|l|q|c|d)^*}$
\\
GenEchare
&$\rm{(b^+|r)^?(m|o^+|f|a^+|l^+|q^+|c|d)^*}$
\\
IntelliJ IDEA
&$\rm{r^?b^*(a|d|o|m|q|c|l|f)^*}$
\\
Trang
&$\rm{b^*(r|(a|d|o|m|q|c|l|f)^+)}$
\\
InstanceToSchema
&$\rm{m^?\&q^*\&b^*\&f^?\&a^*\&o^*\&c^?\&d^?\&r^?\&l^+}$
\\
$learner_{DME}^+$
&$\rm{(q^*|f^?|r^?)\&(o^*|d^?|m^?)\&(a^*|b^*)\&c^*\&l^*}$	
\\
conMiner
&$\rm{r^?b^*c^*o^*d^?m^?f^?\&a^*q^*\&l^*}$	
\\
GenICHARE
&	$\rm{(b^+|r)^?(a^*q^*d^?m^?\&c^+o^*f^?\&l^*)^?}$
\\
GenESIRE
&	$\rm{(b^+|r)^?(a^*(m^?|q^*|d)\&c(o^*|f)\&l^*)^?}$	
\\
$i\mathrm{SOIRE}$
&$\rm{b^*((a^+(q^*|d^?)|m)\&c(o^*|f)\&l^*)|r}$
\\
\hline
\end{tabular}
\label{tab:pos3333}
\end{table}

\begin{table}[htbp]
\sisetup{round-mode=places
        ,round-precision=2
        ,scientific-notation=false
    }
\center
\scriptsize
\renewcommand\arraystretch{2.0}
\caption{Results of inference using different learning algorithms/inference tools on www.}
%\begin{tabular}{c|[1pt]p{5.6cm}<{\centering}}
\begin{tabular}{|p{2.2cm}<{\centering}|p{1.6cm}<{\centering}|p{1.2cm}<{\centering}|p{0.6cm}<{\centering}|p{0.6cm}<{\centering}|}
\hline
\footnotesize{\textbf{Method}}
&\footnotesize{\bm{$|\mathcal{L}(r)|$}}
&\footnotesize{\textbf{DC}}&\footnotesize{\bm{$\mathcal{L}en$}}&\footnotesize{\textbf{ND}}
\\\hline
Liquid Studio
&\num{1.1111E+21}	&$15158.773$	&$120$	&$2$
\\
Soa2Sore
&\num{1.3031E+12}&$7190.139$	&$165$& $3$
\\
Soa2Chare
&\num{1.3553E+19}&$13696.752$&$120$	& $1$
\\
GenEchare
&\num{1.3365E+19}&$13685.703$&$150$	& $2$
\\
IntelliJ IDEA
%&\num{1.1859E+19}&	$13696.313$	&$120$&$1$
&\num{1.3553E+19}&$13696.752$&$120$	& $1$
\\
Trang
&\num{1.2047E+19}&$13606.698$&	$125$&	$1$
\\
InstanceToSchema
&\num{1.5339E+18}&	$13406.824$	&$145$&	$1$
\\
$learner_{DME}^+$
&\num{1.4338E+15}&	$11150.850$&	$175$&$1$	
\\
conMiner
&\num{4.1147E+13}	&$10453.822$&	$145$	&$1$
\\
GenICHARE
&\num{1.4056E+13}	&$9961.492$&	$170$& $2$
\\
GenESIRE
&	\num{4.3899E+11}	&$8479.873$&	$175$& $2$
\\
$\bm{i\mathrm{SOIRE}}$
&\num[math-rm=\bm]{1.83783E+11}&	$\bm{7599.996}$	&$\bm{165}$	&$\bm{1}$
\\
\hline
\end{tabular}
\label{tab:pos4444}
\end{table}

It is clear from the above analysis, our work outperforms other state-of-the-art learning algorithms and published tools, achieving the highest preciseness and the equal level of conciseness. {\color{bella} Furthermore, through the comparison, the performance of our method indicates that the involvement of interleaving could contribute to both preciseness and conciseness. }

\iffalse
In conclusion, even compared with published tools in real world, our work still outweights the performances in terms of preciseness to the large extent, and reaches the same level as the best conciseness. These results show the promissing prospects of our work, as well as the high possibility of application in practical\iffalse the real world\fi.
\fi

\section{Conclusion and Future Work}\label{sec5}

Based on large-scale real data, we proposed a new subclass SOIRE of regular expressions with interleaving. SOIRE is more powerful than the existing subclasses and has unrestricted support for interleaving. Correspondingly, we design an inference algorithm $i$SOIRE which can learn SOIREs effectively based on single occurrence automaton (SOA) and maximum independent set (MIS). We conduct a series of experiments, comparing the performance of our algorithm with both ongoing learning algorithms in academia and industrial tools in real-world.
The results reveal the practicability of SOIRE and the effectiveness of $i$SOIRE, showing the high preciseness and conciseness of our work.

We will study another subclass of regular expressions: $k$-occurrence regular expressions with interleaving ($k$-OIREs) in our future work. Its inference algorithm will also be considered.

\section{ACKNOWLEDGMENTS}

This work is supported by the National Natural Science Foundation of China under Grant Nos. 61872339 and 61472405.
\bibliography{a}
\bibliographystyle{splncs04}

\end{document}